\documentclass[12pt,english,aps,tightenlinesletterpaper,superscriptaddress,secnumarabic,nofootinbib]{revtex4}
\usepackage{mathptmx}

\usepackage[latin9]{inputenc}
\usepackage{babel}
\usepackage{amsmath}
\usepackage{amssymb}
\usepackage{graphicx}
\usepackage[unicode=true,pdfusetitle,
 bookmarks=true,bookmarksnumbered=false,bookmarksopen=false,
 breaklinks=false,pdfborder={0 0 1},backref=section,colorlinks=false]
 {hyperref}
\hypersetup{
 citecolor=blue}

\makeatletter
\@ifundefined{textcolor}{}
{%
 \definecolor{BLACK}{gray}{0}
 \definecolor{WHITE}{gray}{1}
 \definecolor{RED}{rgb}{1,0,0}
 \definecolor{GREEN}{rgb}{0,1,0}
 \definecolor{BLUE}{rgb}{0,0,1}
 \definecolor{CYAN}{cmyk}{1,0,0,0}
 \definecolor{MAGENTA}{cmyk}{0,1,0,0}
 \definecolor{YELLOW}{cmyk}{0,0,1,0}
}

\usepackage{amsfonts}\usepackage{bbm}\usepackage{euscript}
\usepackage{dcolumn}
\usepackage{bm}
\usepackage{epsfig}\epsfclipon
\usepackage{slashed}

\makeatother

\begin{document}

\title{Large Blue Spectral Isocurvature Spectral Index Signals Time-Dependent
Mass}

\author{Daniel J. H. Chung}

\email{danielchung@wisc.edu}

\affiliation{Department of Physics, University of Wisconsin-Madison, Madison,
WI 53706, USA}

\affiliation{Kavli Institute for Cosmological Physics, University of Chicago,
Chicago IL 60637, USA}

\begin{abstract}
We show that if a spectator linear isocurvature dark matter field
degree of freedom has a constant mass through its entire evolution
history, the maximum measurable isocurvature spectral index that is
consistent with the current tensor-to-scalar ratio bound of about
$r\lesssim0.1$ is about $n_{I}\lesssim2.4$, even if experiments can be
sensitive to a $10^{-6}$ contamination of the predominantly adiabatic
power spectrum with an isocurvature power spectrum at the shortest
observable length scales. Hence, any foreseeable future measurement of
a blue isocurvature spectral index larger than $\sim 2.4$ may provide
nontrivial evidence for dynamical degrees of freedom with
time-dependent masses during inflation. The bound is not sensitive to
the details of the reheating scenario and can be made mildly smaller
if $r$ is better constrained in the future.
\end{abstract}
\maketitle

\section{Introduction}

Although minimal single-field slow-roll inflationary
scenarios \cite{Starobinsky:1980te,Sato:1980yn,Linde:1981mu,Mukhanov:1981xt,Albrecht:1982wi,Hawking:1982my,Guth1982,Starobinsky:1982ee,Bardeen:1983qw,Freese:1990rb}
can successfully provide a dynamical explanation for the currently known features of the initial conditions in classical  
cosmological physics (e.g.~the cosmic
microwave background (CMB) \cite{Ade:2015xua,Ade:2015lrj,Ade:2013xsa,Ade:2013lta,Ade:2013rta,Ade:2013sta,Ade:2013ydc,Keisler:2015hfa,Hinshaw:2012fq,Komatsu:2010fb,Brown:2009uy,Reichardt:2008ay,Fowler:2010cy,Lueker:2009rx,Hikage:2012be}
and large scale structure \cite{Ross:2012sx,Percival:2007yw,Eisenstein:2005su}),
it is natural to speculate that more than one single real
field is dynamical during inflation. For such extra dynamical degrees
of freedom not to spoil the flatness of the inflaton potential, it
is also natural to assume that they
are very weakly coupled to the inflaton (though this is obviously not a requirement). With this assumption, these extra dynamical degrees
of freedom behave as spectators as far as the inflationary dynamics is concerned.  If one of these dynamical degrees of freedom is taken to be a weakly interacting
cold dark matter (CDM) field, then there
exists a well-known observable called the CDM-photon isocurvature perturbations
which becomes observable (e.g. \cite{Preskill:1982cy,Abbott:1982af,Dine:1982ah,Axenides:1983hj,Steinhardt:1983ia,Turner:1983sj,Linde:1984ti,Turner:1985si,Seckel:1985tj,Kolb:1990vq,Polarski:1994rz,Gordon:2000hv,Fox:2004kb,Sikivie:2006ni,Beltran:2006sq,Hertzberg:2008wr,Malik:2008im,Langlois:2011zz,Visinelli:2014twa,Choi:2014uaa,Kawasaki:2014una,Harigaya:2015hha,Kadota:2015uia})
if the CDM field is sufficiently weakly interacting and do not to thermalize. 

There are two broad categories of scalar spectator field scenarios that can produce observable CDM-photon
isocurvature perturbations: (i) linear
spectators, such as axions \cite{Peccei:1977hh,Weinberg:1977ma,Wilczek:1977pj},  and (ii) gravitationally produced superheavy dark matter scenarios, aka
WIMPZILLAs \cite{Chung:1998zb,Chung:1998ua,Kuzmin:1999zk,Kuzmin:1998kk,Chung:2001cb,Chung:1998bt}
(for some recent developments, see \cite{Aloisio:2015lva,Fedderke:2014ura,Chung:2013rda,Chung:2013sla}).
Linear spectator fields are characterized by having vacuum expectation
values (VEVs) that are much larger than the amplitudes of their quantum fluctuations.
The VEV oscillations generate the dark matter density in the universe
today while the spatially inhomogeneous distribution of their energy-momentum
tensors are determined by the quantum fluctuations. Such isocurvature fluctuations are called linear because the the energy-momentum
tensor inhomogeneity is approximately linear in the fluctuations,
in contrast with the the case of gravitationally produced superheavy
dark matter scenarios. In this paper, we will focus on the linear
spectator scenarios and will drop the ``linear'' adjective.%
\footnote{We briefly discuss what would happen with
a quadratic isocurvature scenario in the conclusions.%
}

Scale-invariant isocurvature perturbations with negligible correlations
with curvature perturbations are well constrained to be less than
3\% of the adiabatic power \cite{Ade:2015lrj,Ade:2013rta,Hinshaw:2012fq,Komatsu:2010in,Valiviita:2009ck,Sollom:2009vd,Komatsu:2008ex,Bean:2006qz,Crotty:2003fp}.
However, isocurvature spectra with very blue spectral indices can be
unobservably small on long wavelengths, for which the measurements are
strongly constraining, but have large amplitudes on short wavelengths,
where the measurements are less constraining \cite{Hikage:2008sk,Takeuchi:2013hza,Chung:2015pga}.
The case of a blue spectrum is qualitatively different from a ``bump'' in the
spectrum because bumps usually involve a red part as well as a blue
part, and because the blue spectrum here is envisioned to have a qualitatively
extended $k$-space range over which an approximately constant blue
spectral index persists.%
\footnote{Of course, from an observational point of view, this may not be easy
to disentangle since observations have a finite $k$ range.%
} 

One of the most natural models that can produce large blue CDM-photon
isocurvature scenarios was given in \cite{Kasuya:2009up}. This class
of models is characterized by axions that have time-dependent masses
due to the out-of-equilibrium nature of the Peccei-Quinn (PQ) symmetry
breaking field. For constant mass linear spectator fields, large blue-spectral
indices are difficult to produce in observably large amplitudes because
the energy density of the VEV dilutes away. An intuitive perspective
is that the closer the spectral index is to $n_{I}=1$ (scale invariant), the more 
the field fluctuations behave like a frozen VEV, while the closer the spectral
index is to $n_{I}=4$, the more the field fluctuations behave as particles
which can be diluted away by inflation.

Hence, a natural question, which is the subject of this paper, is what is the maximal
measurable isocurvature spectral index that can be produced by a constant
mass spectator field in the context of slow-roll inflationary scenarios
where the adiabatic perturbation spectrum originates from the inflaton field 
fluctuations. For a linear spectator scenario, we find that the maximum
measurable spectral index in the foreseeable future is about $n_{I}=2.4$ (where $n_I=1$ corresponds to scale invariance).
Although measurability depends on the sensitivity of any given experiment,
inflationary physics renders the dependence of the experimental sensitivity
to be logarithmic (to obtain some intuition, see e.g. Eq.~(\ref{eq:maxindex})).
The bulk of this number originates from the ratio of the log of the dark
matter density maximum enhancement due to the dark matter diluting as
$a^{-3}$ (compared to radiation diluting as $a^{-4}$) and the number
of efoldings necessary for the inflationary scenario to explain the observed
homogeneity and isotropy of the universe. A better constraint on the inflationary
tensor perturbation amplitude $r$ can decrease this number, but the
sensitivity is only logarithmic. If restrictions are placed on the maximum
reheating temperature, then the maximum measurable spectral index
also decreases. We will illustrate this by assuming a perturbative reheating
scenario and assuming that the gravitationally suppressed nonrenormalizable
operators of dimension 5 or 6 are unavoidable.

The number 2.4 is interesting because there are claims in the
literature
\cite{Dent:2012ne,Chluba:2013dna,Sekiguchi:2013lma,Takeuchi:2013hza}
that future experiments may be able to measure spectral indices of
$n_{I}\gtrsim3$.  The results in this work demonstrate that if any of
these experiments detect a blue isocurvature spectrum, then they may
have uncovered evidence for a dynamical degree of freedom with a
time-dependent mass.

Before proceeding, we note that the CDM-photon isocurvature observable
that we focus on in this paper is distinct from the $\zeta$ correlator
in the context of ``heavy'' masses discussed e.g.~in
\cite{Arkani-Hamed:2015bza,Chen:2012ge,Craig:2014rta} and the
$\zeta$-tensor correlators \cite{Dimastrogiovanni:2015pla} which in
some cases can also receive signatures from the isocurvature degrees
of freedom. On the other hand, these works all include the common
theme of secondary fields from inflation that can leave a blue
spectral cosmological observable signature.

The order of presentation will be as follows. In Sec.~\ref{sec:Maximization-Problem},
we discuss the constraints considered in the spectral index maximization
problem (there will turn out to be thirteen constraints). We then estimate the solution to the maximization
problem analytically in Sec.~\ref{sec:Analytic-Estimate}. Next, we solve the maximization problem numerically in Sec.~\ref{sec:Numerical-Result}.
We then in Sec.~\ref{sec:What-Happens-With} give a brief review of why the axionic models that naturally have time dependent masses can
evade this bound and explain why this may be the most natural scenario to turn to if measurements are made of the spectral index that are larger than $\sim 2.4$.
Finally, we summarize and discuss caveats in the conclusions.

\section{\label{sec:Maximization-Problem}Maximization Problem}

In this section, we define our class of models and the maximization
problem at hand. In particular, we provide a definition of a measurable blue
isocurvature spectral index for a real scalar field $\chi$ of constant
mass that makes up a fraction $\omega_{\chi}$ of the total
cold dark matter content through its background VEV oscillations, reminiscent
of misaligned axion scenarios.

We consider effectively single-field slow-roll inflationary scenarios,
in which adiabatic cosmological perturbations arise from the inflaton
fluctuations. Here we define effectively single field to mean that a single
field direction is important for the adiabatic inflationary observables.
For example, hybrid inflation involves at least two fields, but during
the slow-roll phase, only one field is dynamical as far as the adiabatic
perturbations are concerned. 

In this context, consider a linear spectator
isocurvature field $\chi$ (see \cite{Chung:2015pga} for a more precise
definition) that is governed by the potential
\begin{equation}
V(\chi)=\frac{m^{2}}{2}\chi^{2},
\end{equation}
in which $m$ is a constant. Writing $\chi=\chi_{0}(t)+\delta\chi(t,\vec{x})$,
the background equation of motion on the metric $ds^{2}=dt^{2}-a^{2}(t)|d\vec{x}|^{2}$
is
\begin{equation}
\partial_{t}^{2}\chi_{0}+3H\partial_{t}\chi_{0}+m^{2}\chi_{0}=0,\label{eq:backgroundeqchi0}
\end{equation}
in which as usual $H\equiv\dot{a}/a$. In accordance with the linear
spectator definition, we assume that for the wave vector $k$ in the range
of isocurvature observable of interest, we have 
\begin{equation}
\chi_{0}(t_{k})\gg\frac{H(t_{k})}{2\pi},
\end{equation}
in which $t_{k}$ is the time when the mode $k$ left the horizon
(i.e. $k=a(t_{k})H(t_{k})$).  The energy density in $\chi_{0}$
oscillations that remains today is assumed to be part of the total
cold dark matter content. We can then divide the $\delta \chi$
(non-inflaton) perturbation into the adiabatic and non-adiabatic part
in the Newtonian gauge as
$\delta\chi_{k}=\delta\chi_{k}^{ad}+\delta\chi_{k}^{nad}$, in which
the nonadiabatic classical isocurvature field fluctuation
$\delta\chi_{k}^{nad}=h_{k}(t)$ obeys the equation
\begin{equation}
\ddot{h}_{k}+3H\dot{h}_{k}+m^{2}h_{k}=0\label{eq:approxmodeeq}
\end{equation}
at the linearized level during inflation. If the approximate Bessel
function solution index $\sqrt{9/4-m^{2}/H^{2}}$ is real while the
modes are subhorizon, then the square root of the $\chi$-photon total
isocurvature amplitude $\sqrt{\Delta_{s}^{2}(k)}$ is 
\begin{equation}
\sqrt{\Delta_{s}^{2}(k)}\approx\omega_{\chi}\frac{\sqrt{k^{3}/(2\pi^{2})}2|h_{k}(t)|}{|\chi_{0}(t)|}
\end{equation}
(we have assumed the usual Bunch-Davies normalization of $h_{k}\rightarrow1/\sqrt{2k}$
in the limit of $k/(aH)\rightarrow\infty$), which remains frozen upon the 
horizon exit, where $\omega_{\chi}$ is the cold dark matter fraction
constituted by $\chi$, assuming all of dark matter is cold. More
precisely, the gauge invariant isocurvature spectrum is (in the notation
of \cite{Chung:2015pga}) 
\begin{equation}
\sqrt{\Delta_{s}^{2}(k)}=\omega_{\chi}2\left(\frac{2^{\nu-\frac{1}{2}}|\Gamma(\nu)|}{\sqrt{\pi}}\right)\left(\frac{H(t_{k_{0}})/(2\pi)}{\chi_{0}(t_{k_{0}})}\right)\left(\frac{k}{k_{0}}\right)^{\frac{3}{2}-\nu+O(\epsilon_{k_{0}})},\label{eq:basicspec}
\end{equation}
in which
\begin{equation}
\nu=\frac{3}{2}\sqrt{1-\frac{4}{9}\frac{m^{2}}{H^{2}(t_{k_{0}})}},
\end{equation}
in which $H(t_{k_{0}})$ is the expansion rate when the $k_{0}$ mode leaves
the horizon. Hence, for the blue spectral indices that are of interest in this work, we have
\begin{equation}
n_{I}-1\approx3-2\nu,\label{eq:specindexandnu}
\end{equation}
in which $\nu$ is a function of mass that also controls the time dependence of the background
field $\chi_{0}$. This class of isocurvature perturbations
will be uncorrelated with the curvature perturbations.

We will call $n_{I}-1\sim O(1)>0$ a large blue spectral index, which  
corresponds to $m/H(t_{k_{0}})\sim O(1)$. For the majority of this paper,
we will take $k_{0}=k_{i}$, which labels the longest wavelength mode
relevant for CMB observations (around $0.002$ Mpc$^{-1}$), and we will 
assume $50\epsilon_{k_{i}}\ll4-n_{I}$. For brevity, we will also
define
$H_{i}\equiv H(t_{k_{i}})$.

As $\omega_{\chi}\propto\chi_{0}^{2}(t)$ in Eq.~(\ref{eq:basicspec}),
and $\chi_{0}(t)$ decays exponentially during inflation whenever
$m/H_{i}\sim O(1)$, $\Delta_{s}^{2}$ can easily become unmeasurably small 
for large blue spectral index scenarios. This suppression
can be partially offset by $(k/k_{0})^{n_{I}-1}$ enhancements as
long as
\begin{equation}
\mbox{constraint 1}:\,\,\,\sqrt{\Delta_{s}^{2}(k_{\rm max})}/\omega_{\chi}<1 \label{eq:constraint1}
\end{equation}
to maintain perturbativity. In addition,  $H_{i}$ cannot in general be made arbitrarily
large to make $\Delta_{s}^{2}$ large due to model building
constraints such as the minimum number of efolds, reheating, and
tensor perturbation limits. 

Given these constraints, a natural question arises:
\begin{itemize}
\item Given an
experimental sensitivity parameterized by $E_{k_{max}}$ (which will be defined below), what is the maximum measurable $n_{I}$ that can be attributed
to a constant mass spectator model in the context of effectively single-field
inflation?
\end{itemize}
 This is the main question that will be answered in this
paper, and the rest of the constraints (together with Eq.~(\ref{eq:constraint1})) associated with maximizing
$n_{I}$ in Eq.~(\ref{eq:basicspec}) will be laid out in this section.

The main physics computation underlying this question is the determination of the time evolution of  
$\chi_{0}(t)$ until the time of reheating. The computation thus depends on
the expansion rate $H(t)$ during and after inflation. More specifically,
the $\chi_{0}(t)$ time dependence is governed by the time-coarse-grained
amplitude of $H$ because Eq.~(\ref{eq:backgroundeqchi0}) does not
contain any derivatives of $H(t)$. To cover a large class of slow-roll
models economically (including both hybrid type and chaotic type),
we consider a coarse-grained model space parameterized by $H_{i}$
(the expansion rate when the the longest wavelength left the horizon),
$\epsilon_{k_{i}}$ (the potential slow-roll parameter when the longest
mode left the horizon), and $t_{e}-t_{k_{i}}$. More precisely, we parameterize
the expansion rate as
\begin{equation}
H\approx\left\{ \begin{array}{ccc}
H_{i}(1-\epsilon_{k_{i}}H_{i}(t-t_{k_{i}})) &  & t_{k_{i}}<t<t_{e}\\
\frac{H_{i}(1-\epsilon_{k_{i}}H_{i}(t_{e}-t_{k_{i}}))}{1+\frac{3}{2}(t-t_{e})H_{i}(1-H_{i}\epsilon_{k_{i}}(t_{e}-t_{k_{i}}))} &  & t>t_{e}
\end{array}\right.\label{eq:generalfittingfunction}
\end{equation}
which is continuous at $t_{e}$.%
\footnote{Since we will never take the derivative of this function at
  $t_{e}$ in the computation, the discontinuity of the derivative at
  $t=t_{e}$ does not pose significant inaccuracies for the spectator
  field. This expansion rate fits the quadratic inflationary model to
  better than 10\% during most of the time except at the inflationary
  exit transition where the fit degrades to 40\% accuracy briefly at
  the transition point out of quasi-dS era. For inflationary models
  with smaller $\epsilon_{k_{i}}$, the fit is better, since this is a
  perturbative solution in $\epsilon_{k_{i}}$.  An alternative to this
  approach would be a numerical $H$ time evolution sampling in the
  space of single field slow-roll models
  \cite{Lidsey:1995np,Kinney:2002qn,Kinney:2003uw,Chung:2005hn,Agarwal:2008ah,Powell:2008bi}.
  We do not invest in the more numerically intensive approach since
  even an order 40\% uncertainty in $H$ amounts to an order 1\%
  uncertainty in $n_{I}-1$ in most of the parametric regime of
  interest.  More discussion of this  will be given later.  } We will consider
$\epsilon_{k_{i}}$ values that are consistent with the single-field
adiabatic perturbation amplitude
\begin{equation}
\mbox{constraint 2:}\,\,\,\,\,\,\epsilon_{k_{i}}(H_{i})=\frac{H_{i}^{2}}{8\pi^{2}M_{p}^{2}\Delta_{\zeta}^{2}(k_{i})},\label{eq:adiabaticconstraint}
\end{equation}
in accordance with the spectator isocurvature paradigm considered in
this paper. In the above, $\Delta_{\zeta}^{2}(k_{i})$ is the adiabatic spectral
amplitude at the longest observable wavelengths, which we will take
to be $\Delta_{\zeta}^{2}(k_{i})\approx2.4\times10^{-9}$.%
\footnote{This is consistent with current Planck measurements \cite{Ade:2015xua}.
A 10\% change in this number only leads to less than a 1\% change in
our results, while we are aiming for a 10\% accuracy in $n_{I}-1$.
Hence, the precision of this number is not very important.%
} As we seek a conservative upper bound on 
$n_{I}-1$, we will not impose the adiabatic scalar
spectral index constraint.%
\footnote{The imposition of the adiabatic spectral index constraint using a
full chain of slow-roll parameter evolution scenarios will not give
a severe constraint on $t_{e}$ because of the large functional degree
of freedom that exists in the inflationary slow-roll potential space,
and its inclusion will obscure the presentation needlessly.%
} The set of models that this parameterization excludes are those for which the quantities 
$\{\epsilon_{k_{i}},\, H_{i},\, t_{e}\}$ do not control $H(t_{e})$,
the expansion rate at the end of inflation. Such excluded models are
somewhat atypical among known set of explicit effectively single-field
models as they require new length scales (i.e. beyond $H_{i}$ and
$t_{e}$) to enter the potential beyond those that are typically present in hybrid and
chaotic inflationary scenarios. Furthermore, new length scales
require yet another degree of fine tuning to fit smoothly with the $t\sim t_{k_{i}}$
time region where Eq.\ (\ref{eq:generalfittingfunction}) is guaranteed
to be valid for effectively single-field slow-roll models. As we will
discuss later, the maximum spectral index constraint does not sensitively
depend on $\epsilon_{k_{i}}H_{i}t_{e}$, which is fortunate since
this parameterization is only 40\% accurate for quadratic inflationary
models near the time of the end of inflation. Note also that because
we will impose the tensor perturbation phenomenological upper bound
on $H_{i}$, the $\epsilon_{k_{i}}$ contribution to the spectral index
will never be too big for phenomenological compatibility.

In addition to the adiabatic constraint Eq.~(\ref{eq:adiabaticconstraint}),
we impose the inflationary condition that the number of efolds be
larger than the minimum necessary for a successful cosmology:
\begin{eqnarray}
\mbox{constraint 3:\,\,\,}N_{e}\equiv H_{i}\Delta t_{e}\left[1-\frac{\epsilon_{k_{i}}}{2}H_{i}\Delta t_{e}\right]>N_{min} & \approx & 53+\frac{1}{3}\ln\frac{T_{RH}}{10^{10}{\rm GeV}}-\frac{2}{3}\ln\frac{H_{e}(H_{i},t_{e})}{10^{10}{\rm GeV}},\label{eq:nefoldmin-1}
\end{eqnarray}
in which
\begin{equation}
H_{e}\equiv H_{i}(1-\epsilon_{k_{i}}H_{i}(\Delta t_{e})), \qquad
\Delta t_{e}\equiv t_{e}-t_{k_{i}},
\end{equation}
and we have taken the largest length scale to be $k_{{\rm min}}\sim2\pi H_{0}a_{0}$.
Note that in writing Eq.~(\ref{eq:nefoldmin-1}), we are neglecting contributions of order
$\ln\left(c_{h}H_{i}/(2H_{e})\right)$, in which $c_{h}$ is an inflationary
model dependent function of order unity. This leads to a systematic
uncertainty with approximately a 2\% error in the isocurvature spectral index
bound. Note also Eq.~(\ref{eq:nefoldmin-1}) is a non-linear constraint
on $H_{i}$.

We also impose the constraint that arises from assuming that there is
at least one gravitational strength operator that can reheat the
universe. Such assumptions are well motivated within string-motivated
cosmologies
(e.g.~\cite{Blumenhagen:2014gta,Cicoli:2012cy,Cicoli:2010ha,Green:2007gs,Chen:2006ni,Chialva:2005zy,Frey:2005jk,Kofman:2005yz})
and the weak gravity conjecture \cite{ArkaniHamed:2006dz} (for some
recent developments, see
e.g.~\cite{Brown:2015iha,Rudelius:2015xta,Cheung:2014vva}), as well
  as generic expectations of interpreting gravity as an effective
  theory with the cutoff scale $M_{p}$. The minimum reheat temperature
  for a given $H_{e}$ can be computed assuming a coherent oscillation
  perturbative reheating. For the inflaton field degree of freedom
  $\varphi$ at the end of inflation to oscillate, we must have its
  mass $m_{\varphi}$ satisfy the condition $m_{\varphi}\gtrsim
  H_{e}$. If the particle decay is through a dimension
  $n_{\mathcal{O}}\geq5$ operator, then
\begin{equation}
\Gamma_{g}\sim S \frac{m_{\varphi}^{2(n_{\mathcal{O}}-4)+1}}{M_{p}^{2(n_{\mathcal{O}}-4)}}
\end{equation}
is the gravitational decay rate representing the ``weakest'' decay
rate where $ S $ is a phase space suppression factor. For
2-body decay, we expects $ S \sim(8\pi)^{-1}$, and we will
take $ S $ as small as $(0.1)^{2}/(8\pi)$ to get a conservative
bound. Since
\begin{equation}
T_{RH}=0.2\left(\frac{200}{g_{*}(T_{RH})}\right)^{1/4}\sqrt{\Gamma M_{p}},
\end{equation}
in which $\Gamma$ is the total decay rate, the bound $\Gamma\gtrsim\Gamma_{g}$
and $m_{\varphi}\gtrsim H_{e}$ lead to the following bound
\begin{equation}
\mbox{constraint 4:\,\,\,\,\,\,}H_{e}\lesssim H_{e\mbox{ rh bound}}(T_{RH})\equiv\left[\left(\frac{T_{RH}}{0.2\left(\frac{200}{g_{*}(T_{RH})}\right)^{1/4}}\right)^{2}\frac{M_{p}^{2(n_{\mathcal{O}}-4)-1}}{ S }\right]^{\frac{1}{2(n_{\mathcal{O}}-4)+1}}.\label{eq:Heupperbd-1}
\end{equation}
As we will see, for the maximal spectral index bounds at the highest
reheating temperatures, this constraint is unimportant. A further constraint
from reheating is that $\{H_{i},\epsilon_{k_{i}},t_{e}\}$ has to
be chosen for a fixed reheating temperature such that the energy at
the end of inflation is large enough to give the total radiation energy:
\begin{equation}
\mbox{constraint 5:\,\,\,\,\,\,}T_{RH}<\left(\frac{10}{g_{*}}\right)^{1/4}\sqrt{\frac{3}{\pi}M_{P}H_{e}}.
\end{equation}
Here we have implicitly assumed $T_{RH}$ and $H_{e}$ are such that
coherent oscillations of $\chi$ occurs during the oscillation period
of the inflaton. This condition can be written as 
\begin{equation}
\frac{3}{2}H(t_{RH})<m,
\end{equation}
which can be used to put a lower bound on the spectral index of
\begin{equation}
\mbox{constraint 6:\,\,\,\,\,\,}n_{I}-1>3-\sqrt{9-\left[\frac{\pi^{2}}{10}g_{*}(T_{RH})\left(\frac{T_{RH}^{2}}{M_{p}H_{i}}\right)^{2}\right]}.\label{eq:bound-1}
\end{equation}
Although imposing constraint 6 seems artificial since it 
is a simplification for calculational and presentation purposes, the
parametric region where this bound is relevant is very similar to
the parametric region where constraint 5 is relevant (i.e.,~it excludes
the similar $\{H_{i},t_{e}\}$ region). Hence, there is no qualitative
change in computing the maximum $n_{I}-1$. Furthermore, we find in
the explicit numerical work that $n_{I}-1$ bound is lowered through
constraint 6 by less than 1\% which is below the systematic uncertainty
in the computation. Hence, constraint 6 is a posteriori not important
as long as constraint 5 is imposed.

The absence of observed tensor perturbations yield the following phenomenological
bound:
\begin{equation}
\mbox{constraint 7:\,\,\,\,\,\,}H_{i}<M_{p}\sqrt{\frac{r_{b}}{2}\Delta_{\zeta}^{2}(k_{i})}
\end{equation}
in which $r_{b}$ is the bound on the tensor-to-scalar ratio (i.e. the
ratio $r<r_{b} \simeq 16\epsilon_{k_{i}}$).
For the dark matter fraction to not exceed unity, we impose another
phenomenological bound of
\begin{equation}
\mbox{constraint 8:}\,\,\,\,\,\,\,\,\omega_{\chi}\leq1.
\end{equation}
We see that constraints $2-5$ and 7 mainly arise from inflationary model-building consistency, while constraint 8 deals with dark matter phenomenology.

We now turn to constraints on isocurvature perturbations in addition
to constraints 1 and 6. Let us suppose that future experiments can detect isocurvature
amplitudes $\sqrt{\Delta_{s\, o}^{2}(k_{\rm max})}$ above $E_{k_{\rm max}}\sqrt{\Delta_{\zeta}^{2}(k_{i})}$, in which
$E_{k_{\rm max}}$ parameterizes the experimental sensitivity.
Eq.~(\ref{eq:basicspec}) implies
\begin{equation}
\mbox{constraint 9: }\,\,\,\,\,\,\,\omega_{\chi}2\left(\frac{2^{\nu-\frac{1}{2}}|\Gamma(\nu)|}{\sqrt{\pi}}\right)\left(\frac{H_{i}/(2\pi)}{\chi_{0}(t_{k_{i}})}\right)\left(\frac{k_{\rm max}}{k_{i}}\right)^{\frac{3}{2}-\nu}\geq E_{k_{\rm max}}\sqrt{\Delta_{\zeta}^{2}(k_{i})},
\end{equation}
in which we have assumed $3/2-\nu\gg\epsilon_{k_{i}}$. We note that neglecting 
$\epsilon_{k_{i}}$ in the spectral index is numerically valid to
better than 2\% level for the upper bound of interest. 

To see that constraint 9 controls the bound on the isocurvature $n_{I}-1$
that we are seeking, we note that if $\chi_{0}$ oscillations occur before
reheating, we have
\begin{equation}
\omega_{\chi}=\frac{m^{2}\langle\left(\chi_{0}\right)^{2}\rangle_{t=t_{RH}}\left(\frac{a(t_{RH})}{a(t_{eq})}\right)^{3}}{\rho_{R}(T_{eq})(\Omega_{DM}/(\Omega_{b}+\Omega_{DM}))},\label{eq:origfrac-1}
\end{equation}
in which $\Omega_{DM}$ is the total dark matter fraction of the critical
density today, $\Omega_{b}$ is the total baryonic fraction today,
and $t_{eq}$ is the time of matter-radiation equality. The prediction
from the coherent oscillation perturbative reheating scenario takes the form
\begin{equation}
\omega_{\chi}=R\frac{2\langle\left(\chi_{0}\right)^{2}\rangle_{t=t_{RH}}}{M_{p}^{2}}\left[\frac{m}{H(t_{RH})}\right]^{2}\left(\frac{T_{RH}}{T_{eq}}\right),
\end{equation}
in which
\begin{equation}
  R\equiv\frac{\Omega_{b}+\Omega_{DM}}{\Omega_{DM}}\frac{1}{6}\frac{g_{*}(T_{RH})}{g_{*}(T_{eq})}\frac{g_{*S}(T_{eq})}{g_{*S}(T_{RH})}\approx\frac{\Omega_{b}+\Omega_{DM}}{\Omega_{DM}}\frac{3.94}{3.38}\frac{1}{6}\approx0.23,
  \label{eq:Rdef}
\end{equation}
where $g_{*}(T)$ counts the degrees of freedom in the radiation energy
density $\rho_{R}$, $g_{*S}(T)$ counts the degrees of freedom in
the entropy density, and $T_{eq}\approx0.8$ eV is the matter-radiation
equality temperature.%
\footnote{Here we used $g_{*S}(T_{eq})=3.94$ and $g_{*}(T_{eq})=3.38$.%
} As the solution to Eq.~(\ref{eq:backgroundeqchi0}) is approximately given by
\begin{equation}
\left(\chi_{0}(t)/\chi_{0}(t_{k_{i}})\right)^{2}\sim e^{-(n_{I}-1)H_{i}(t-t_{k_{i}})}
\end{equation}
during inflation, while the next most important factor is 
\begin{equation}
\left(\frac{k}{k_{i}}\right)^{\frac{3}{2}-\nu}=e^{\frac{n_{I}-1}{2}\ln\left(\frac{k}{k_{i}}\right)}
\end{equation}
with $\ln[k/k_{i}]\ll$ (number of efolds of inflation), we see that
the magnitude of the left hand side of constraint 9 is controlled
by $\omega_{\chi}$ and will be monotonically decreasing as $(n_{I}-1)/2$
increases in the blue spectral parametric region of interest.%
\footnote{We see that intuitively when $n_{I}-1=0$, the background field acts
like a time independent constant while when $n_{I}-1\rightarrow3^{-}$,
the field behaves as a diluting gas of non-relativistic particles. %
} Hence, we conclude that the maximum $n_{I}-1$ is obtained when we
\emph{saturate} the inequality of constraint 9.

It is also necessary to check the current phenomenological bound on the isocurvature perturbations:
\begin{equation}
\sqrt{\frac{\Delta_{s}^{2}(k_{1})}{\Delta_{\zeta}^{2}(k_{1})}}<E_{k_{1}},\label{eq:const10precursor}
\end{equation}
in which the current phenomenological bound on $E_{k_{1}}$ for $k_{1}\approx0.05$
Mpc$^{-1}$ is $\sim 0.2$ at 95\% confidence level \cite{Ade:2015lrj}.
Since $\Delta_{s}^{2}(k)\propto k^{n_{I}-1}$, when constraint 9 is saturated Eq.~(\ref{eq:const10precursor}) becomes
\begin{equation}
\frac{E_{k_{\rm max}}\sqrt{\Delta_{\zeta}^{2}(k_{i})}\left(\frac{k_{1}}{k_{\rm max}}\right)^{\frac{n_{I}-1}{2}}}{\sqrt{\Delta_{\zeta}^{2}(k_{1})}}<E_{k_{1}}.
\end{equation}
To simplify the approximate phenomenological constraint parameterization,
we choose $k_{1}=k_{i}$: 
\begin{equation}
\mbox{constraint 10}:\,\,\,\,\,\,\, E_{k_{\rm max}}\left(\frac{k_{i}}{k_{\rm max}}\right)^{\frac{n_{I}-1}{2}}<E_{k_{i}}.
\end{equation}
Finally, we must also make sure we are in the linear spectator regime
with our choice of $k_{\rm max}$:
\begin{equation}
\mbox{constraint 11:}\,\,\,\,\,\,\,\,\chi_{0}(t_{k_{\rm max}})>\frac{H(t_{k_{\rm max}})}{2\pi}
\end{equation}
and $\chi_{0}(t_{k_{i}})$ is not trans-Planckian:
\begin{equation}
\mbox{constraint 12:}\,\,\,\,\,\,\,\,\chi_{0}(t_{k_{i}})\leq M_{p}.
\end{equation}
There is another uncertainty in constraint 9 that is associated with
the fact that the Bessel function mode functions are not obviously
accurate solutions whenever the slow-roll parameter is not negligible.
The limitations due to this issue were spelled out in \cite{Chung:2015pga}.
A more accurate power-law expansion should have a fiducial value
of $k_{0}=k_{\rm max}$ instead of $k_{i}$. (The price that is paid for
doing this is a complicated/numerical expression for
$\chi_{0}(t_{k_{\rm max}})$ in terms of $\chi_{0}(t_{k_{i}})\leq M_{p}$.)
This will turn out to be an issue only for values of $H_{i}$ that saturate
constraint 7 with $r_{b}\gg10^{-2}$ because $m/H(t)$ evolves significantly
in that case during inflation. To address this issue, for such worrisome situations we therefore check the following constraint numerically
\begin{equation}
\mbox{constraint 9': }\omega_{\chi}\frac{\sqrt{k^{3}/(2\pi^{2})}2|h_{k}(t_{k_{i}})|}{|\chi_{0}(t_{k_{i}})|}\geq E_{k_{\rm max}}\sqrt{\Delta_{\zeta}^{2}(k_{i})}
\end{equation}
involving a more accurate set of numerical solutions only.  Finally, we note that constraint 9 also assumes that
\begin{equation}
\mbox{constraint 13: \,\,\,\,\,\,\,\,\,\,\,}\frac{m}{H(t_{k_{\rm max}})}<\frac{3}{2},
\end{equation}
since only the non-decaying mode has been kept. We will see that in practice this
does not pose a significant constraint. 

In summary, the problem of finding the maximally observable constant
mass isocurvature spectral index $n_{I}$ for a given experimental
sensitivity $E_{k}$ is to find the maximum $n_{I}$ that satisfies the constraints 1-13 given above.

\section{\label{sec:Analytic-Estimate}Analytic Estimate}

In this section, we provide an analytical estimate of the solution to the $n_{I}-1$
extremization problem presented in Sec.~\ref{sec:Maximization-Problem}. We begin in 
Section \ref{sub:Without-Slow-roll-Evolution} by giving a crude estimate
of the maximization problem that is obtained by neglecting the slow-roll parameter
$\epsilon_{k_{i}}$. In Section \ref{sub:Perturbative-in-Slow-roll}, we then obtain an analytic perspective of the effect of turning
on the slow-roll evolution of $H$ and the non-linearities of the
problem. For example, we will see that the $H_{i}$ may not quite
saturate constraint 7 for the largest spectral index, in contrast
with the estimate given in Section \ref{sub:Without-Slow-roll-Evolution},
and this turns out to be significant for the accuracy of the approximation
of the spectral index used in constraint 9. Sec.~\ref{sec:Numerical-Result} will involve a numerical solution to the constrained maximization
problem without resorting to the analytic arguments presented in this section.

\subsection{\label{sub:Without-Slow-roll-Evolution}Without Slow-roll Evolution}

As previously discussed, the maximal spectral index results when constraint 9 is saturated.  To evaluate constraint 9, we need to determine $\langle\left(\chi_{0}\right)^{2}\rangle_{t=t_{RH}}$.
In this section, we will estimate this quantity to obtain a qualitative understanding
of the parameters involved. 

Let us neglect the slow-roll evolution
of $H$ and assume that $\chi_{0}$ coherently oscillates
just at the end of inflation. We can then estimate
\begin{equation}
\langle\left(\chi_{0}\right)^{2}\rangle_{t=t_{RH}}\sim C\left[\frac{H(t_{RH})}{H(t_{e})}\right]^{2}\label{eq:decreaseamplitude}
\end{equation}
\begin{equation}
C\equiv\frac{1}{2}\chi_{0}^{2}(t_{k_{i}})\exp\left[-(n_{I}-1)N_{e}\right],
\end{equation}
in which $N_{e}\approx H_{i}(t_{e}-t_{k_{i}})$ is the number of efolds
of inflation. Through standard cosmological scaling, this yields the
dark matter fraction to be
\begin{equation}
\omega_{\chi}\sim R\frac{\chi_{0}^{2}(t_{k_{i}})\exp\left[-(n_{I}-1)N_{e}\right]\left[\frac{m}{H_{i}}\right]^{2}}{M_{p}^{2}}\left(\frac{T_{RH}}{T_{eq}}\right).
\end{equation}
Now, noting that $m/H_{i}\sim O(1)$, and that the greatest $n_{I}-1$ sensitivity
comes from the exponential, we find (assuming constraint 1 is satisfied)
\begin{eqnarray}
n_{I}-1 & \lesssim & \frac{55}{N_{e}\left(1-\frac{1}{2N_{e}}\ln\left[\frac{k_{max}}{k_{i}}\right]\right)}\left(1+\frac{1}{55}\ln\left[\frac{10^{-5}2\left(\frac{2^{\nu-\frac{1}{2}}|\Gamma(\nu)|}{\sqrt{\pi}}\right)\left(\frac{H_{i}/(2\pi)}{\chi_{0}(t_{k_{i}})}\right)}{E_{k_{\rm max}}\sqrt{\Delta_{\zeta}^{2}(k_{i})}}\left(\frac{T_{RH}}{10^{10}{\rm GeV}}\right)\right]\right.\nonumber \\
 &  & \left.+\frac{1}{55}\ln\left[\frac{\left(\chi_{0}\right)_{t_{k_{min}}}^{2}R}{M_{p}^{2}}\right]\right),\label{eq:zerothordercond9sol}
\end{eqnarray}
in which we note that $\nu=(3-[n_{I}-1])/2$. Hence, we see that increasing $T_{RH}$, $k_{max}/k_{i}$, and $H_{i}$ while decreasing
$N_{e}$ and $E_{k_{max}}$ is what we want to maximize $n_{I}$.
Clearly, $N_{e}$ cannot be decreased beyond the minimal number of efolds $N_{min}$ that is 
necessary for a successful inflationary scenario (constraint 3) for a fixed $T_{RH}.$
This will be one of the strongest constraints for bounding
$n_{I}-1$. Increasing $H_{i}$ while keeping $N_{e}$ (and $T_{RH}$)
fixed requires decreasing $t_{e}$, since $N_{e}\sim H_{i}(t_{e}-t_{k_{i}})$.
However, because $N_{min}$ also changes if $H_{i}$ and $t_{e}$
changes, it is not possible to keep $N_{e}$ fixed right at the constraint
boundary. As $H_{i}$ keeps increasing, it eventually runs into the
tensor perturbation constraint 7. Also relevant for the case of low reheating
temperatures is the fact that for sufficiently large $H_{i}/T_{RH}$,
we run into constraint 4. For each $T_{RH}$, 
$n_{I}-1$ can be maximized through the $H_{i}$ and $t_{e}$ variations subject to
the constraints just described.

As $T_{RH}$ is increased towards the highest temperatures consistent
with energy conservation, constraints 5 and 6 become relevant. Even
though constraint 6 is a tiny bit stronger of a constraint, it is
very similar in numerical value to constraint 5. This is fortunate because
as described before, constraint 6 is imposed for computational convenience
and constraint 5 arises from fundamental principle of energy conservation.
In the $\{H_{i},t_{e}\}$ parametric region where constraints 5 and
6 compete, the reheating scenario is somewhat unrealistic in that
the reheating time scale is very fast, taking the system away from
the coherent oscillation perturbative reheating regime. However, to
put a conservative upper bound, we account for this extreme parametric
region as well. It is in this sense that the bound that we will obtain
for the maximum $n_{I}$ is reheating scenario independent.

We next note that for lower reheating temperatures satisfying constraint
7 
\begin{equation}
H_{e\mbox{ rh bound}}(T_{RH})<M_{p}\sqrt{\frac{r}{2}\Delta_{\zeta}^{2}}
\end{equation}
(with $H_{i}$ maximized to maximize $n_{I}-1$), $H_{i}$ has to be
brought down when $T_{RH}$ is brought down to satisfy constraint 4:
\begin{equation}
H_{i}\sim\left[\left(\frac{T_{RH}}{0.2\left(\frac{200}{g_{*}(T_{RH})}\right)^{1/4}}\right)^{2}\frac{M_{p}^{2(n_{\mathcal{O}}-4)-1}}{ S }\right]^{\frac{1}{2(n_{\mathcal{O}}-4)+1}}.
\end{equation}
Since we have saturated constraint 9, we see that 
\begin{equation}
\omega_{\chi}=\frac{E_{k_{\rm max}}\sqrt{\Delta_{\zeta}^{2}(k_{i})}}{2\left(\frac{2^{\nu-\frac{1}{2}}|\Gamma(\nu)|}{\sqrt{\pi}}\right)\left(\frac{H_{i}/(2\pi)}{\chi_{0}(t_{k_{i}})}\right)\left(\frac{k_{max}}{k_{i}}\right)^{\frac{3}{2}-\nu}}
\end{equation}
increases as $H_{i}$ is lowered. Hence, depending in particular on
the numerical values of $E_{k_{\rm max}}$ and $k_{max}/k_{i}$,
the required $\omega_{\chi}$ can exceed unity, violating constraint
8. 

Finally, if we choose $k_{i}/k_{\rm max}\lesssim10^{-5}$ and $E_{k_{i}}/E_{k_{\rm max}}\gtrsim3\times10^{-3}$,
we can always satisfy constraint 10 if $n_{I}-1\sim1$. The current
scale invariant isocurvature perturbation bound is given by
\begin{equation}
\frac{\Delta_{s}^{2}(100k_{i})}{\Delta_{\zeta}^{2}(100k_{i})}\lesssim3\times10^{-2}.
\end{equation}
 If this scale invariant spectrum bound is assumed to bound the blue
spectrum as well, we have 
\begin{equation}
E_{k_{i}}\sim\sqrt{3\times10^{-2}\times10^{-2}}\sim10^{-2}
\end{equation}
 for $n_{I}-1\sim1$. Hence, we see that if we choose $E_{k_{\rm max}}\lesssim1$
and $k_{i}/k_{\rm max}\lesssim10^{-5}$ , we can satisfy constraint
10. Therefore, we will focus on this parametric regime and ignore constraint
10.

Let us now find an explicit estimate of the largest value of $n_{I}$. First, we consider
the case of 
\begin{equation}
T_{RH}>T_{n_{\mathcal{O}}}
\end{equation}
\begin{equation}
T_{n_{\mathcal{O}}}\equiv2^{\frac{5}{2}-\frac{n_{\mathcal{O}}}{2}}\left(\Delta_{\zeta}^{2}\right)^{\frac{n_{\mathcal{O}}}{2}-\frac{7}{4}}M_{p}r^{-\frac{7}{4}+\frac{n_{\mathcal{O}}}{2}}\sqrt{\frac{S}{5\sqrt{g_{*}(T_{RH})}}}\approx\begin{cases}
2\times10^{10}r^{3/4}\sqrt{S8\pi}{\rm GeV} & \,\,\,\,\,\,\,\,\, n_{\mathcal{O}}=5\\
6.8\times10^{5}r^{5/4}\sqrt{S8\pi}{\rm GeV} & \,\,\,\,\,\,\,\,\, n_{\mathcal{O}}=6,
\end{cases}
\end{equation}
for which constraint 7 becomes relevant. Saturating constraints
3 and 7 in the current approximation scheme, we find
\begin{equation}
N_{e}\sim52.8+\frac{1}{3}\ln\frac{T_{RH}}{10^{10}{\rm GeV}}-\frac{2}{3}\ln\frac{M_{p}\sqrt{\frac{r}{2}\Delta_{\zeta}^{2}}}{10^{10}{\rm GeV}}.
\end{equation}
Although $T_{RH}$ appears here suggesting $T_{RH}$ should be minimized
to maximize the $n_{I}-1$ bound, the $T_{RH}$ dependence shown explicitly
in Eq.~(\ref{eq:zerothordercond9sol}) dominates. As constraints
5 and 6 are similar in magnitude, we use constraint 5 to maximize
$T_{RH}$ for simplicity for this simplified analytic estimate. In
other words, here we estimate
\begin{eqnarray}
\mbox{max }T_{RH} & \approx & \left(\frac{10}{g_{*}}\right)^{1/4}M_{p}\sqrt{\frac{3}{\pi}\sqrt{\frac{r}{2}\Delta_{\zeta}^{2}}}\\
 & \sim & 6.5\times10^{15}r^{1/4}\mbox{ GeV },\label{eq:estimatemax}
\end{eqnarray}
in which we have taken $g_{*}=200$ and $\Delta_{\zeta}^{2}(k_{i})=2.4\times10^{-9}$.
From Eq.~(\ref{eq:zerothordercond9sol}), we then find that
\begin{eqnarray}
n_{I}-1|_{T_{RH}=\mbox{max}T_{RH}} & \lesssim & \frac{55}{N_{e}^{est}\left(1-\frac{1}{2N_{e}^{est}}\ln\left[\frac{k_{max}}{k_{i}}\right]\right)}\left(1+\frac{1}{55}\ln\left[\frac{\left(\chi_{0}\right)_{t_{k_{min}}}^{2}R}{M_{p}^{2}}\right]\right.\nonumber \\
 &  & \left.\frac{1}{55}\ln\left[\frac{4.2\times10^{3}\left(\frac{2^{\nu-\frac{5}{4}}|\Gamma(\nu)|}{\pi^{2}}\right)\left(\frac{M_{p}}{\chi_{0}(t_{k_{i}})}\right)r^{3/4}}{E_{k_{\rm max}}}\left(\frac{10}{g_{*}}\Delta_{\zeta}^{2}(k_{i})\right)^{1/4}\right]\right)\label{eq:precursortomaxindex}\\
 & \approx & \boxed{\frac{1.2\times\left(1+\frac{1}{54}\ln\left[\frac{r^{3/4}}{E_{k_{\rm max}}}\right]\right)}{1-5.5\times10^{-3}\ln r-1.1\times10^{-2}\ln\left[\frac{k_{max}}{k_{i}}10^{-5}\right]}},\label{eq:maxindex}
\end{eqnarray}
in which we have taken $g_{*}$ and $\Delta_{\zeta}^{2}(k_{i})$ to have the same numerical values as above. We have also taken $\chi_{0}(t_{k_{i}})=M_{p}$ to
satisfy constraint 12, 
we have 
approximated $n_{I}-1\approx1$ on the right hand side, and we used
\begin{equation}
N_{e}^{est}=52.8+\frac{1}{3}\ln\frac{\left(\frac{10}{g_{*}}\right)^{1/4}M_{p}\sqrt{\frac{3}{\pi}\sqrt{\frac{r}{2}\Delta_{\zeta}^{2}(k_{i})}}}{10^{10}{\rm GeV}}-\frac{2}{3}\ln\frac{M_{p}\sqrt{\frac{r}{2}\Delta_{\zeta}^{2}(k_{i})}}{10^{10}{\rm GeV}}\approx51.2\left(1-\frac{\ln r}{205}\right).
\end{equation}
We see that there is only a very mild dependence
on the phenomenological parameterizations used above. Hence, if a blue isocurvature
spectral index $n_{I}\gtrsim3$ is measured, this certainly cannot arise from a linear spectator with a time-independent mass. We will
sharpen this estimate with a numerical analysis in Sec.~\ref{sec:Numerical-Result}.

We next consider the case of
\begin{equation}
T_{n_{\mathcal{O}}}\lesssim T_{RH}\lesssim\left(\frac{10}{g_{*}}\right)^{1/4}M_{p}\sqrt{\frac{3}{\pi}\sqrt{\frac{r}{2}\Delta_{\zeta}^{2}(k_{i})}}\approx6.5\times10^{15}r^{1/4}{\rm GeV}
\end{equation}
 but still with $H_{i}$ saturating constraint 7:
\begin{equation}
\boxed{\begin{array}{ccc}
n_{I}-1 & \lesssim & \frac{1.06}{1-8\times10^{-3}\ln r+8\times10^{-3}\ln\frac{T_{RH}}{10^{11}{\rm GeV}}-1.2\times10^{-2}\ln\left[\frac{k_{max}}{k_{i}}10^{-5}\right]}\times\\
 &  & \left(1+\frac{1}{44}\ln\left[\frac{\sqrt{r}}{E_{k_{\rm max}}}\right]+\frac{1}{44}\ln\frac{T_{RH}}{10^{11}{\rm GeV}}\right)
\end{array}}\label{eq:noindependentslope-1}
\end{equation}
where we have used
\begin{equation}
N_{e}^{est}\approx47.5-\frac{1}{3}\ln r+\frac{1}{3}\ln\frac{T_{RH}}{10^{11}{\rm GeV}}.
\end{equation}
We can continue to lower the temperature towards $T_{n_{\mathcal{O}}}$ unless
constraint 8 is saturated. Constraint 8 is saturated before reaching
$T_{n_{\mathcal{O}}}$ if $n_{I}-1$ is smaller than the solution
$n_{I}^{c}(T>T_{n_{\mathcal{O}}})-1$ to
\begin{equation}
\frac{2^{\frac{n_{I}^{c}-1}{2}-1}\chi_{0}(t_{k_{i}})\sqrt{\Delta_{\zeta}^{2}(k_{i})}E_{k_{\rm max}}\left(\frac{k_{\rm max}}{k_{{\rm min}}}\right)^{-\frac{(n_{I}^{c}-1)}{2}}\pi^{3/2}}{\kappa\sqrt{r}\Gamma\left(\frac{3-(n_{I}^{c}-1)}{2}\right)}=1,\label{eq:definitionofnic}
\end{equation}
which for $\{E_{k_{\mbox{max }}}=1,k_{\rm max}/k_{\mbox{min }}=10^{5},\chi_{0}(t_{k_{i}})=M_{p},r_{b}=10^{-3}\}$
is approximately $0.7$.

Let us now consider the lower reheating temperature $T_{RH}<T_{n_{\mathcal{O}}}$ (and $n_{I}-1>n_{I}^{c}-1$),
which means that we should set
\begin{equation}
H_{i}\approx H_{e\mbox{ rh bound}}(T_{RH})\label{eq:belowthebreak}
\end{equation}
 in Eq.~(\ref{eq:zerothordercond9sol}) instead of using constraint
7. We find
\begin{equation}
\boxed{n_{I}-1\lesssim\begin{cases}
\frac{1.06\left(1-0.023\ln\left[E_{k_{\rm max}}\right]+0.038\ln\frac{T_{RH}}{10^{11}{\rm GeV}}-7.6\times10^{-3}\ln\left[S8\pi\right]\right)}{1+5.4\times10^{-3}\ln\left[S8\pi\right]-2.7\times10^{-3}\ln\frac{T_{RH}}{10^{11}{\rm GeV}}-1.2\times10^{-2}\ln\left[\frac{k_{max}}{k_{i}}10^{-5}\right]} & \,\,\,\,\,\,\,\,\, n_{\mathcal{O}}=5\\
\frac{1.2\left(1-0.021\ln\left[E_{k_{\rm max}}\right]+0.030\ln\frac{T_{RH}}{10^{11}{\rm GeV}}-4.2\times10^{-3}\ln\left[S8\pi\right]\right)}{1+3.5\times10^{-3}\ln\left[S8\pi\right]+1.7\times10^{-3}\ln\frac{T_{RH}}{10^{11}{\rm GeV}}-1.3\times10^{-2}\ln\left[\frac{k_{max}}{k_{i}}10^{-5}\right]} & \,\,\,\,\,\,\,\,\, n_{\mathcal{O}}=6
\end{cases}}\label{eq:nodependentTrhdependence}
\end{equation}
 At lower $T_{RH}$, we have $\omega_{\chi}\gtrsim1$ at 
\begin{equation}
T_{DM2}=\begin{cases}
\frac{4.6\times10^{7}\mbox{ GeV }E_{k_{\rm max}}^{3/2}\sqrt{8\pi S}}{\left(\frac{k_{\rm max}/k_{{\rm min}}}{10^{5}}\right)^{3/4}}\left(1-(n_{I}-2)\left[8.5+\frac{3}{4}\ln\left(\frac{k_{\rm max}/k_{{\rm min}}}{10^{5}}\right)\right]\right) & \,\,\,\,\,\,\,\,\,\,\,\,\,\,\,\,\, n_{\mathcal{O}}=5\\
\frac{28\mbox{ GeV }E_{k_{\rm max}}^{5/2}\sqrt{8\pi S}}{\left(\frac{k_{\rm max}/k_{{\rm min}}}{10^{5}}\right)^{5/4}}\left(1-(n_{I}-2)\left[14+\frac{5}{4}\ln\left(\frac{k_{\rm max}/k_{{\rm min}}}{10^{5}}\right)\right]\right) & \,\,\,\,\,\,\,\,\,\,\,\,\,\,\,\,\, n_{\mathcal{O}}=6.
\end{cases}
\end{equation}
We have taken the minimum $T_{RH}$ in this paper to be at 100 GeV
to simplify the presentation. This means that $\omega_{\chi}\lesssim1$
constraint is more relevant for $n_{\mathcal{O}}=5$ case than the
$n_{\mathcal{O}}=6$ case.

We should also estimate the effect of constraint 11 on $k_{\rm max}$:
\begin{equation}
\frac{k_{\rm max}}{k_{i}}\lesssim\left(\frac{\chi_{0}(t_{k_{i}})}{H_{i}/2\pi}\right)^{\frac{2}{n_{I}-1}},
\end{equation}
which becomes
\begin{equation}
\frac{k_{\rm max}}{k_{i}}\lesssim10^{8}\label{eq:kmaxovkminbound}
\end{equation}
 with $\chi_{0}(t_{k_{i}})=M_{p},$ $n_{I}-1=1.3$, and $H_{i}$ set
at $r_{b}=0.1$.

Finally, constraint 13 can be shown to be generically satisfied in
the $n_{I}-1$ and $k_{\rm max}/k_{\mbox{min }}$ region of interest. This will be discussed more in the numerical section below.

\subsection{\label{sub:Perturbative-in-Slow-roll}Perturbative in Slow-roll Evolution}

In this subsection, we examine the effect of turning on $\epsilon_{k_{i}}$.
We will see its most important feature is to have $n_{I}$ maximized
for $|H_{i}-H_{e}|/H_e \ll 1 $, making the Bessel spectral formula accurate.

Instead of completely neglecting $\epsilon_{k_{i}}$ during inflation
in computing $\langle\left(\chi_{0}\right)^{2}\rangle_{t=t_{RH}},$
we can use linear perturbation theory in $\epsilon_{k_{i}}$ to solve
Eq.~(\ref{eq:backgroundeqchi0}) (for more details, see appendix \ref{sec:Background-Sol}
):
\begin{equation}
\langle\left(\chi_{0}\right)^{2}\rangle_{t\geq t_{m},t_{e}}\sim\frac{\mathcal{A}}{2}\chi_{0}^{2}(t_{k_{i}})\exp\left[-(3-2\nu_{k_{i}})H_{i}(t_{e}-t_{k_{i}})\right]\left[\frac{H(t)}{H(t_{e})}\right]^{2}\label{eq:ampsq}
\end{equation}
\begin{equation}
\mathcal{A}\equiv\left(1-\frac{6(3-2\nu_{k_{i}})F_{1}\epsilon_{k_{i}}H_{i}\nu_{k_{i}}(t_{e}-t_{k_{i}})}{16\nu_{k_{i}}^{3}-3(3-2\nu_{k_{i}})\epsilon_{k_{i}}}\right)^{2}\left(1+F_{2}^{2}\right)\label{eq:mathcala}
\end{equation}
\begin{equation}
F_{1}\equiv H_{i}\nu_{k_{i}}(t_{e}-t_{k_{i}})-1
\end{equation}
\begin{equation}
F_{2}\equiv\frac{32\nu_{k_{i}}^{4}+6\epsilon_{k_{i}}\nu_{k_{i}}\left\{ 3-2\nu_{k_{i}}\left[1+H_{i}(t_{e}-t_{k_{i}})\left(3+\nu_{k_{i}}[2+(3-2\nu_{k_{i}})H_{i}(t_{e}-t_{k_{i}})]\right)\right]\right\} +F_{3}}{\sqrt{9-4\nu_{k_{i}}^{2}}\left(16\nu_{k_{i}}^{3}-3\epsilon_{k_{i}}(3-2\nu_{k_{i}})(1+2F_{1}H_{i}\nu_{k_{i}}(t_{e}-t_{k_{i}})\right)}
\end{equation}
\begin{equation}
F_{3}\equiv9(3-2\nu_{k_{i}})\epsilon_{k_{i}}^{2}H_{i}(t_{e}-t_{k_{i}})[1+2F_{1}H_{i}\nu_{k_{i}}(t_{e}-t_{k_{i}})],
\end{equation}
in which $t_{m}$ is the time after $\chi$ field starts to oscillate
and $t_{e}$ is the end of inflation:
\begin{equation}
m=\frac{3}{2}H(t_{m}).
\end{equation}
This allows us to rewrite the analog of Eq.~(\ref{eq:zerothordercond9sol})
as 
\begin{equation}
n_{I}-1\lesssim\frac{55}{N_{min}'}\left(1+\frac{1}{55}\ln\left[\frac{10^{-5}}{\omega_{\chi\,{\rm min}}(n_{I}-1)}\left(\frac{T_{RH}}{10^{10}{\rm GeV}}\right)\right]+\frac{1}{55}\ln\left[\frac{\left(\chi_{0}\right)_{t_{k_{min}}}^{2}r_{m}^{2}(n_{I}-1)R}{M_{p}^{2}}\right]\right),
\end{equation}
in which
\begin{equation}
N_{min}'\approx\frac{1-\sqrt{1-2\epsilon_{k_{i}}N_{min}}}{\epsilon_{k_{i}}}
\end{equation}
\begin{equation}
\omega_{\chi\,{\rm min}}(n_{I}-1)\equiv\mbox{max}\left[E_{k_{\rm max}}\sqrt{\Delta_{\zeta}^{2}(k_{min})},\frac{E_{k_{\rm max}}\sqrt{\Delta_{\zeta}^{2}(k_{min})}}{\sqrt{\Delta_{s}^{2}(k_{\rm max})}/\omega_{\chi}}\right]
\end{equation}
\begin{equation}
r_{m}(n_{I}-1)\equiv\frac{m}{H_{i}(1-\epsilon_{k_{i}}N_{min}')}\sqrt{\mathcal{A}},
\end{equation}
$N_{min}$ is given by Eq.~(\ref{eq:nefoldmin-1}) $\mathcal{A}$
is given by Eq.~(\ref{eq:mathcala}), $\sqrt{\Delta_{s}^{2}(k_{\rm max})}/\omega_{\chi}$
is given by Eq.~(\ref{eq:basicspec}), and
\begin{equation}
\frac{m}{H_{i}}\approx\frac{1}{2}\sqrt{(n_{I}-1)(6-[n_{I}-1])}.
\end{equation}
Comparing with Eq.~(\ref{eq:zerothordercond9sol}), we see a complicated
function $r_{m}(n_{I}-1)$ that depends on $n_{I}-1$. Most of this
complicated function accounts for the $\epsilon_{k_{i}}$ dependence
of the time evolution of $\chi_{0}(t)$. 

When accounting for $\epsilon_{k_{i}}$ and constraint 2, we note that a given
pair $H_{e}$ and $N_{e}$ can originate from two different values $H_{i}$:
\begin{equation}
H_{i}=\frac{\sqrt{\Delta_{\zeta}^{2}}M_{p}\sqrt{2\pi}\sqrt{\pi\pm\sqrt{\pi^{2}-\frac{H_{e}^{2}N_{e}}{M_{p}^{2}\Delta_{\zeta}^{2}}}}}{\sqrt{N_{e}}}.
\end{equation}
With $H_{e}^{2}N_{e}\ll\pi^{2}M_{p}^{2}\Delta_{\zeta}^{2}$, the hybrid
inflation case corresponds to the minus sign branch while the quadratic
inflation case corresponds to the positive sign branch. One can also easily
show that for the parametric regime of interest, $H_{e}$ never
becomes close to zero even though there may be a worry from the form of Eq.~(\ref{eq:generalfittingfunction})
that we may be unreasonably extrapolating the linear expansion of
the slow-roll that is valid near $t_{k_{i}}$. On the other hand,
the parametric regions where the hybrid inflation and quadratic inflation
branches merge are sensitive to the branchpoint singularity there.

The most important feature of turning on $\epsilon_{k_{i}}$ is that
since now (with constraint 2 imposed)
\begin{equation}
N_{e}=\frac{4\pi^{2}\Delta_{\zeta}^{2}M_{p}^{2}}{H_{i}^{2}}\left(\frac{H_{i}^{2}}{H_{e}^{2}}-1\right),\label{eq:nonsaturation}
\end{equation}
the minimization of $N_{e}$ that is important for the extremization of $n_{I}-1$
(e.g.~see Eq.~(\ref{eq:zerothordercond9sol}) which in turn is related
to constraint 9) gives a numerical pressure in the non-linear extremization
problem to make $H_{i}$ close to $H_{e}$. This favors a smaller
$\epsilon_{k_{i}}$ (in turn favoring small $H_{i}$) which competes
with the pressure to extremize $H_{e}$ (favoring a large $H_{i}$)
that arises from constraint 3. Hence, depending on the size of $r$, $H_{i}$
may not quite saturate constraint 7 as was done in the derivation
of Eq.~(\ref{eq:maxindex}). This means that with $\epsilon_{k_{i}}$
turned on, the sensitivity to the tensor-to-scalar ratio $r$ entering
constraint 7 is reduced for values of $r$ that are ``large''. As we will see
in Sec.~\ref{sec:Numerical-Result}, this makes the approximate spectral
index Eq.~(\ref{eq:specindexandnu}) more accurate for $r_{b}=0.1$.

A figure of validity for the $\epsilon_{k_{i}}$ perturbations can be written
as 
\begin{equation}
C_{pert}\equiv\epsilon_{k_{i}}N_{e}^{2},
\end{equation}
since the background field evolution equation during inflation 
\begin{equation}
\partial_{t}^{2}\chi_{0}+3H_{i}(1-\epsilon_{k_{i}}H_{i}(t-t_{k_{i}}))\partial_{t}\chi_{0}+m^{2}\chi_{0}=0
\end{equation}
has a secular term $\epsilon_{k}H_{i}(t-t_{k_{i}})$, and this term
is integrated over a time period of $t_{e}-t_{k_{i}}\propto N_{e}$.
Since $N_{e}\sim50$, high $H_{i}$ models where $\epsilon_{i}$ approaches
the tensor-to-scalar ratio $r$ bound have $C_{pert}$ approaching
unity, and hence they cannot be addressed reliably using this perturbative approach. In the next section, we will turn to a numerical analysis of this extremization problem, which will allow us to get a handle on situations such as these when perturbative methods fail.

\section{\label{sec:Numerical-Result}Numerical Results}

\begin{figure}
\begin{centering}
\includegraphics[scale=0.6]{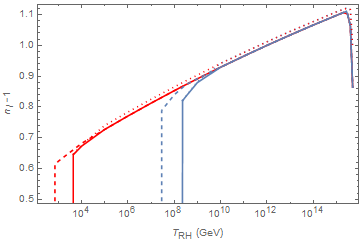}~~~~~\includegraphics[scale=0.6]{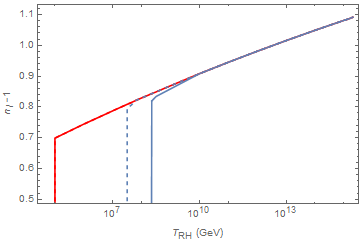}
\par\end{centering}

\protect\caption{\label{fig:maxspectralindex-1}The maximum measurable spectral index
$n_{I}-1$ assuming $\{E_{k_{\rm max}}=1,\, k_{\rm max}/k_{{\rm min}}=10^{5};\, r_{b}=10^{-1},10^{-3};\, n_{\mathcal{O}}=5,6;\, S=(8\pi)^{-1},10^{-2}(8\pi)^{-1}\}$
is plotted as a function of $T_{RH}$. (left) The bound on the tensor
to scalar ratio $r_{b}$ has been taken to be $10^{-1}$. The maximum
spectral index with the amplitude parameterized by $\{E_{k_{\rm max}}=1,\, k_{\rm max}/k_{{\rm min}}=10^{5}\}$
is around $n_{I}=2.11$. For $T\lesssim10^{9}$ GeV, the lower solid
curve corresponds to the reheating non-nonrenormalizable operator dimension
of $n_{\mathcal{O}}=5$ while the upper solid curve corresponds to
the case of the non-nonrenormalizable decay operator dimension of $n_{\mathcal{O}}=6$.
For $T\gtrsim10^{9}$ GeV, the two curves merge. The dashed curve
corresponds to weakening the coefficient of the non-nonrenormalizable
operator by a factor of 10. The vertical curve on the left portion
of the boundary curves occur because the expansion rate there is too
small in that parametric regime to produce measurable isocurvature
perturbations (i.e. constraint 8). The dotted curve corresponds to
evaluation of fiducial spectral index at $k_{\rm max}$ instead
of $k_{{\rm min}}$. The correction is small (except at the highest $T_{RH}$ where the dip occurs) because constraint 7
is not saturated for $r_{b}=0.1$. (right) Similar to the left plot
except with a probably possible future bound of $r_{b}=10^{-3}$. }
\end{figure}

In this section, we perform a numerical analysis to find the largest $n_{I}$ consistent with constraints
1 through 13.   The results of this analysis will show that even with an extremely optimistic experimental sensitivity of $10^{-6}\Delta_{\zeta}^{2}$
on length scales as small as 10 kpc scales, the theoretical prediction from a constant mass isocurvature field scenario is that experiments
will not measure spectral indices $n_{I}$ greater than $2.4$.

We begin with Fig.~\ref{fig:maxspectralindex-1}, which shows the case in which $\{E_{k_{\rm max}}=1,\, k_{\rm max}/k_{{\rm min}}=10^{5};\, r_{b}=10^{-1},10^{-3};\, n_{\mathcal{O}}=5,6;\, S=(8\pi)^{-1},10^{-2}(8\pi)^{-1}\}$.
The results show that the maximum temperature estimated in Eq.~(\ref{eq:estimatemax})
agrees with the right end of each plot to better than 30\% and the
maximum $n_{I}-1$ agrees with Eq.~(\ref{eq:maxindex}) to better
than 5\%. For the $r_{b}=0.1$ plot (the left plot), the reason why there
is a drop of $n_{I}-1$ near $T_{RH}\sim5\times10^{15}$ GeV is due
to constraint 5 (reheating energy conservation at time $t_{e}$) pushing
up $H_{i}$ as $T_{RH}$ is raised.%
\footnote{This increases $\epsilon_{k_{i}}$, which in turn increases the split between
$H_{i}$ and $H_{e}$. This then increases $N_{e}$, as can be seen in Eq.~(\ref{eq:nonsaturation}),
under the assumption that the increase in the split is the most important
effect. %
} This upward push of $H_{i}$ is allowed because from the discussion
around Eq.~(\ref{eq:nonsaturation}), constraint 7 may not be
saturated depending on the size of $r.$ This non-saturation is indeed
the case for most of the $r_{b}=0.1$ curve (which we have also checked
directly numerically) and makes the approximate spectral index Eq.~(\ref{eq:specindexandnu})
more accurate.  We see how the dotted curve matches the solid curve except
at the highest temperature where the dip occurs, as we will discuss more below.%
\footnote{\label{fn:nonimportanceofsliver}The bottom of the dip is where the
mismatch of the accurate dotted curve and the approximate solid curve
is the largest. This does not affect our main result since it does
not correspond to globally the largest spectral index. Furthermore,
this reheating sliver is where the reheating scenario is least realistic
and has been considered only to give a conservative bound on $n_{I}$.%
} For the $r_{b}=10^{-3}$ case, constraint
7 does saturate at the highest allowed reheating temperature, which
means that no upward push of $H_{i}$ ever arises from constraint
5 for these highest temperatures. The maximum $n_{I}-1$ for this
$\{E_{k_{\rm max}}=1,\, k_{\rm max}/k_{{\rm min}}=10^{5}\}$
experimental scenario is about 1.1. Any measurements of CDM-photon
blue isocurvature with a spectral index larger than $n_{I}=2.1$ with
an amplitude larger than equal to $\{E_{k_{\rm max}}=1,\, k_{\rm max}/k_{{\rm min}}=10^{5}\}$
imply the responsible dynamical degree of freedom during inflation
cannot be a constant mass linear spectator field.

Let us now consider some of the other features of these results. For all but one of
the curves shown in Fig.~\ref{fig:maxspectralindex-1}, when $T_{RH}$
from above to below $T_{n_{\mathcal{O}}}$ while $n_{I}-1>n_{I}^{c}-1$
(see Eq.~(\ref{eq:definitionofnic}) for the definition), there is
a break in the bound curve as expected from Eq.~(\ref{eq:belowthebreak})
encoding the minimal reheating constraint 4. The break in the curve
does not exist for the case of $\{r_{b}=10^{-3},\, n_{\mathcal{O}}=6\}$
because in that case $n_{I}-1$ reaches $n_{I}^{c}(T>T_{n_{\mathcal{O}}=6})-1$,
in which
\begin{equation}
T_{n_{\mathcal{O}}=6}(r_{b}=10^{-3},S=(8\pi)^{-1})\approx120{\rm GeV},
\end{equation}
which means that the dark matter constraint 8  is saturated without saturating
the reheating constraint 4. All of the curves
terminate at a certain lower endpoint of reheating temperature because
of the dark matter constraint 8, which simply states that the expansion
rate in that parameter regime is too small to produce measurable isocurvature
perturbations. We note that there is no vertical line plotted for the
right hand side of the curves in Fig.~\ref{fig:maxspectralindex-1}
(unlike the left vertical line) because we did not want to obscure
the drop in $n_{I}-1$ for high $T_{RH}$ for $r_{b}=0.1$.

\begin{figure}
\begin{centering}
\includegraphics[scale=0.6]{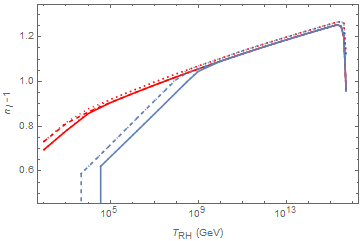}~~~\includegraphics[scale=0.6]{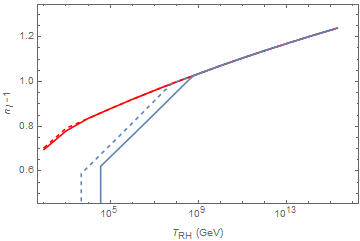}
\par\end{centering}

\protect\caption{\label{fig:maxspectralindexbetterexperiment-1}The maximum measurable
spectral index $n_{I}-1$ as a function of $T_{RH}\in[100,5\times10^{15}]$
GeV assuming an experimental sensitivity of $E_{k_{\rm max}}=10^{-3}$
corresponding to resolving $\Delta_{s}^{2}/\Delta_{\zeta}^{2}$ to
$O(10^{-4}\%)$ at about $1$ Mpc length scale. The rest of the parameters
are set at $\{k_{\rm max}/k_{\rm min}=10^{5};\, r_{b}=10^{-1},10^{-3};\, n_{\mathcal{O}}=5,6;\, S=(8\pi)^{-1},10^{-2}(8\pi)^{-1}\}$.
(left) The bound on the tensor-to-scalar ratio $r$ has been taken
to be $10^{-1}$. The lower solid curve for $T_{RH}\lesssim10^{9}$
GeV corresponds to the reheating non-nonrenormalizable operator dimension
of $n_{\mathcal{O}}=5$ just as in Fig.~\ref{fig:maxspectralindex-1}.
The dotted curve corresponds to evaluation of fiducial spectral index
at $k_{\rm max}$ instead of $k_{{\rm min}}$. As in Fig.~\ref{fig:maxspectralindex-1},
the correction is small because constraint 7 is not saturated even
for $r_{b}=0.1$.  As expected from increasing the experimental resolution
by $10^{3}$, the maximum measurable spectral index has only gone
up mildly to $n_{I}=2.25$ (from $n_{I}=2.12$). Note that unlike
in Fig.~\ref{fig:maxspectralindex-1}, the bounds for $n_{\mathcal{O}}=6$
end at $T_{RH}=10^{2}$ GeV because we simply truncated the plot there
(and not because $\omega_{\chi}>1$ there). The dashed curve corresponds
to weakening the coefficient of the non-nonrenormalizable operator by
a factor of 10 just as in Fig.~\ref{fig:maxspectralindex-1}. (right)
Similar to the left plot except that $r_{b}$ has been set to $10^{-3}$.}
\end{figure}

Fig.~\ref{fig:maxspectralindexbetterexperiment-1} shows the case
with $\{E_{k_{\rm max}}=10^{-3},\, k_{\rm max}/k_{\rm min}=10^{5};\, r_{b}=10^{-1},10^{-3};\, n_{\mathcal{O}}=5,6;\, S=(8\pi)^{-1},10^{-2}(8\pi)^{-1}\}$.
Decreasing $E_{k_{\rm max}}$ to $10^{-3}$ means increasing the
experimental sensitivity (i.e., $\Delta_{s}^{2}/\Delta_{\zeta}^{2}$ is resolved 
to $10^{-6}$ instead of order unity -- an extremely optimistic view
of the foreseeable future that is chosen to illustrate the insensitivity of
the bound to experimental precision). This changes the measurable
maximal spectral index logarithmically to about $n_{I}=2.25$ (from
$n_{I}=2.1$ when $E_{k_{\rm max}}=1$). Hence, although increasing
experimental sensitivity changes the measurable blue spectral index,
the logarithmic nature of the increase makes these numbers experimentally
meaningful for at least a many decades time scale. As before, the maximum
temperature estimated in Eq.~(\ref{eq:estimatemax}) agrees with
the right end of the plot to better than 30\% and the maximum $n_{I}-1$
agrees with Eq.~(\ref{eq:maxindex}) to better than 10\%. For the
$r_{b}=0.1$ plot (left plot), the reason why there is a drop of $n_{I}-1$
near $T_{RH}\sim5\times10^{15}$ GeV is the same reason as in the
explanation for Fig.~\ref{fig:maxspectralindex-1}. Note that unlike
in Fig.~\ref{fig:maxspectralindex-1}, the bounds for $n_{\mathcal{O}}=6$
end at $T_{RH}=10^{2}$ GeV because we simply truncated the plot there
(and not because $\omega_{\chi}>1$ there). 
\begin{figure}
\begin{centering}
\includegraphics{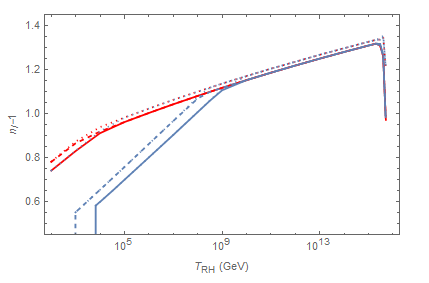}
\par\end{centering}

\protect\caption{\label{fig:largerkmax-1}Similar to the left Fig.~\ref{fig:maxspectralindexbetterexperiment-1}
except with $k_{\rm max}/k_{\rm min}=10^{7}$ (i.e. $k_{\rm max}$
is at the scale of 10 kpc). The bound on the maximum spectral index
$n_{I}$ is only logarithmically sensitive to $k_{\rm max}/k_{\rm min}$
as it is now about 2.35 instead of 2.25. The features of the plots
are explained as in previous figures. As before, the dotted curve
corresponds to the evaluation of the spectral index with the fiducial
$k$ value of $k_{\rm max}$ instead of $k_{\rm min}$.}
\end{figure}

Finally, to be extremely optimistic regarding short distance scale
probes of cosmology, in Fig.~\ref{fig:largerkmax-1} we consider the $n_{I}-1$ bound with an experimental
probe length scale of $k_{\rm max}/k_{\rm min}=10^{7}$ (i.e.
$k_{\rm max}$ is at the scale of 10 kpc) with the other parameters
set at $\{E_{k_{\rm max}}=10^{-3},\, r_{b}=10^{-1},10^{-3};\, n_{\mathcal{O}}=5,6;\, S=(8\pi)^{-1},10^{-2}(8\pi)^{-1}\}$. The maximum spectral index increases
as expected in a mild manner to $n_{I}=2.35$ (from $n_{I}=2.25$
with $k_{\rm max}/k_{\rm min}=10^{5}$). Note that this $k_{\rm max}/k_{\rm min}$
lies near the edge of constraint 11 in accordance with Eq.~(\ref{eq:kmaxovkminbound}).

\begin{figure}
\begin{centering}
\includegraphics[scale=0.6]{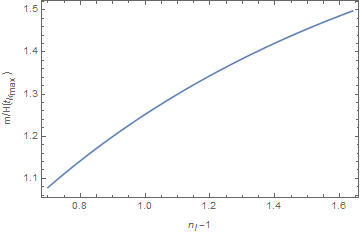}
\par\end{centering}

\protect\caption{\label{fig:rlargeaccurate}$m/H(t_{k_{\rm max}})$ is plotted
as a function of $n_{I}-1$ defined according to Eq.~(\ref{eq:specindexandnu})
for the severest parametric choices of $r_{b}=0.1$ and $k_{\rm max}/k_{\rm min}=10^{7}$.
This shows constraint 13 is satisfied for $n_{I}-1\lesssim1.6$.}
\end{figure}

Also, constraint 13 can be shown to be generically satisfied in the
$n_{I}-1$ and $k_{\rm max}/k_{\mbox{min }}$ region of our interest.
For example, Fig.~\ref{fig:rlargeaccurate} shows $m/H(t_{k_{\rm max}})$
as a function of $n_{I}-1$ defined according to Eq.~(\ref{eq:specindexandnu})
for the parametric choices of $r_{b}=0.1$ and $k_{\rm max}/k_{{\rm min}}=10^{7}$
(which is the most constrained among the scenarios we are interested
in). We see that since we have considered only $n_{I}-1\lesssim1.6$,
constraint 13 will be satisfied.

Note that for the numerical computations discussed thus far, only the
background fields are evolved fully numerically to determine $\omega_{\chi}$
while analytic approximations relying on $H(t)$ being constant have
been used to compute $\Delta_{s}^{2}/\omega_{\chi}^{2}$ in accordance
with \cite{Chung:2015pga}. For $r\ll0.1$ we have $\epsilon_{k_{i}}\ll1$,
and $H(t)$ evolution does not present much of a correction. However,
for $r_{b}=0.1$ in the plots above, there may be a worry that the numerical
computation of $\Delta_{s}^{2}/\omega_{\chi}^{2}$ would deviate significantly
from the approximations. One symptom of the analytic mode functions
destroying the accuracy of $\Delta_{s}^{2}/\omega_{\chi}^{2}$ can
be tested by comparing the answers for two different fiducial values
of $k_{0}$.

The parametric spectral indices $n_{I}-1$ shown in Figs.~\ref{fig:maxspectralindex-1}
through \ref{fig:largerkmax-1} (except for the dotted curves) correspond
to the $(k/k_{0}){}^{n_{I}-1}$ approximate parameterization with $k_{0}$
chosen at $k_{0}=k_{i}$ which is the longest observable wave vector.
For $H_{i}$ corresponding to saturating constraint 7 with $r\lesssim10^{-2}$,
this is a good parameterization: i.e., in Figs.~\ref{fig:maxspectralindex-1}
through \ref{fig:largerkmax-1}, plots with $r_{b}=10^{-3}$ can be
taken to be accurate to better than 1\%. However, \textbf{if }$H_{i}$
saturates the limit of constraint 7 with $r\approx10^{-1}$, $H(t)$
would evolve nontrivially during inflation. In that case, there is a worry 
as to whether the $(k/k_{i})^{n_{I}-1}$ parameterization is inaccurate
for the $r_{b}=0.1$ cases. For example, if we saturate constraint
7 with $r_{b}=10^{-1}$, a more accurate approximation of the observed
spectrum near $k_{\rm max}$ should have the fiducial value $k_{0}=k_{\rm max}$
(at the expense of computing $\chi_{0}(t_{k_{0}})$ numerically).
Fortunately, we find numerically that constraint 7 is never saturated
even with $r_{b}=0.1$ because of the effects discussed in Eq.~(\ref{eq:nonsaturation}).
The accuracy of the analytic spectrum calculation can also be seen
in the dotted curves of Figs.~\ref{fig:maxspectralindex-1} through
\ref{fig:largerkmax-1} which were computed numerically.%
\footnote{ The agreement between the dotted curve and the solid curve exists
except at the highest $T_{RH}$ dipping sliver which does not correspond
to the globally maximum $n_{I}$, as we discussed in footnote \ref{fn:nonimportanceofsliver}.%
}By explicit computation, we have checked that $n_{I}-1$ computed
with mode function evolution evolved fully numerically matches the
$n_{I}-1$ computed through the Bessel function with $k_{0}$ shifted
to $k_{\rm max}$ (and $\chi_{0}(t_{k_{\rm max}})$ computed
numerically) to better than a few percent.

Hence, we conclude from Fig.~\ref{fig:rlargeaccurate} that any measurement
of $n_{I}>2.4$ for CDM-photon isocurvature perturbations in the foreseeable
future indicates the responsible dynamical degree of freedom during
inflation cannot be a constant mass linear spectator field.

\section{\label{sec:What-Happens-With}Models: What Happens With a Dynamical Mass?}

In \cite{Chung:2015pga}, it was shown that spectral indices as large
as $n_{I}=3.8$ (but not $n_{I}=4$) can be achieved in the context of a
dynamical VEV breaking the Peccei-Quinn (PQ) symmetry. This is of
interest because $n_{I}=3$ is considered to be observable for example
by the Square Kilometer Array \cite{Takeuchi:2013hza}. Here, we
discuss why a time-dependent mass during inflation can evade the bound
discussed around Fig.~\ref{fig:largerkmax-1}.  Suppose the mass of the
field $\chi$ responsible for the linear spectator isocurvature makes a
transition at time $t_{c}$ from value $m$ to zero. According to
corollary 2 of \cite{Chung:2015pga}, the modes $k<k_{c}$ that leave
the horizon earlier than the time of the mass transition still have
the form of Eq.~(\ref{eq:basicspec}), where if the slow-roll evolution
is neglected, the critical wave vector is given by
\begin{equation}
k_{c}\sim k_{\rm min}\exp[N_{c}],
\end{equation}
in which
\begin{equation}
N_{c}\equiv H_{i}(t_{c}-t_{k_{\rm min}})
\end{equation}
is the number of efolds from the beginning of inflation. For $k_{c}\sim10^{7}\, k_{\rm min}$,
the number of efolds for which this occurs must be at least $N_{c}\sim16$
efolds. All of these modes are governed by massive scalar field quantum
fluctuations giving a blue spectrum. In addition, the background field $\chi_{0}(t)$
dilutes as
\begin{equation}
\chi_{0}(t)\propto\exp[-\frac{n_{I}-1}{2}N_{c}],
\end{equation}
which dilutes the total isocurvature by an important factor: 
\begin{equation}
\Delta_{s}^{2}\propto\exp[-2\left(n_{I}-1\right)N_{c},]
\end{equation}
which is analogous to Eq.~(\ref{eq:decreaseamplitude}). The field
theory up to this point behaves just as in the constant mass scenarios
we have been discussing. 

However, after the mass transition to masslessness completes, the
background field $\chi_{0}(t)$ behaves as a constant massless field
until the end of inflation. Hence, compared to the constant mass case,
the isocurvature perturbations receive a boost of
\begin{equation}
\frac{\Delta_{s}^{2}(\mbox{time dependent mass})}{\Delta_{s}^{2}(\mbox{constant mass})}\propto\exp[2\left(n_{I}-1\right)(N_{e}-N_{c})]\label{eq:enhancement}
\end{equation}
in which $N_{e}$ is the total number of efolds as usual. Since $N_{e}\sim50$, the enhancement for $N_{c}\sim16$ scenario is enormous. This
is the intuitive explanation with which time-dependent mass situations
can evade the blue spectral index bounds for the time-independent
mass situation that has been the main focus of this paper. One observational
signature of the mass transition \cite{Kasuya:2009up} is the existence
of a flat isocurvature spectrum (for $k>k_{c}$) in addition to the
blue spectrum ($k<k_{c}$). On the other hand, if there is a limited
$k$-range accessible experimentally, it may not be easy to observe
the break in the spectrum.

A natural question is then what class of models naturally produce
these time dependent masses. Note that the crucial ingredient in being
able to generate the large enhancement Eq.~(\ref{eq:enhancement})
is the transition from $m/H_{i}\sim O(1)$ to $m/H_{i}\ll1$. If $H_{i}$
is the natural minimum energy scale for the masses of the scalar dynamical
degrees of freedom (as is the case for example in supergravity models), then a symmetry
needs to naturally lead to $m/H_{i}\ll1$. Hence, one crucial ingredient
for natural isocurvature models with $n_{I}$ larger than the bound
presented for the constant mass case is a symmetry $X$ protecting
the $\chi$ mass from Hubble scale corrections to its mass. A second
ingredient is a temporary (but lasting many efolds) mass generation
mechanism. This second ingredient is necessary to generate the blue
spectrum.

In the supersymmetric axion scenario of \cite{Kasuya:2009up}, the
symmetry $X$ is the Peccei-Quinn (PQ) symmetry non-linearly realized
as a shift symmetry of the axion field. The PQ symmetry breaking fields
$\Phi_{\pm}$ are displaced from the minimum of the effective potential
during inflation (in a way in which PQ symmetry is always broken)
such that the coset symmetry $X$ is actually broken by $\partial_{t}\Phi_{\pm}$
through the kinetic structure of the axion: i.e., the Nambu-Goldstone theorem
does not apply because the system is not in vacuum. As $\Phi_{\pm}$
fields roll toward the vacuum (where the PQ breaking persists), the axions
behave as a massive field with mass of the order of $H_{i}$ due to
the supergravity structure of the K\"{a}hler potential. After $\Phi_{\pm}$
reaches the vacuum and the kinetic energy dilutes to the point of
$\partial_{t}\Phi_{\pm}\ll H_{i}\Phi_{\pm}$, $X$ is restored axions
become massless, up to the small explicit PQ breaking contribution.

Although it is possible to tune parameters and initial conditions to obtain
almost flat potentials, the Nambu-Goldstone models with out-of-equilibrium
symmetry-breaking time-dependent VEVs seem to be the simplest natural
model. From this perspective, any experiment measuring a CDM-photon
isocurvature perturbations with $n_{I}\gtrsim2.4$ may be finding
evidence for a dynamical degree of freedom during inflation that has
a coset shift symmetry.

\section{\label{sec:Conclusions}Conclusions}

We have considered a constant mass spectator linear isocurvature degree
of freedom during inflation and answered the question of what is the largest
measurable blue spectral index that can be produced via such a mechanism. We have shown that the largest measurable spectral index
is less than 2.4 in the foreseeable future with only logarithmic sensitivity
to experimental precision characterized by $\{E_{k_{\rm max}},\, k_{\rm max}/k_{\rm min}\}$
and experimental constraints such as the tensor-to-scalar ratio $r_{b}$.
This means that any future measurements of the isocurvature spectral
index above this bound would give weight to the hypothesis that there
is a spectator field with a time-dependent mass during inflation.

We have also considered how for reheating temperatures much smaller than
the maximum allowed by tensor perturbation bound, the maximum observable
spectral index decreases. This would be relevant if there were specific
inflationary models under consideration with a fixed reheating scenario
or model-dependent phenomenological bounds on the reheating temperature
such as those that arise from cosmologically dangerous gravitinos. For
part of this smaller reheating temperature dependent bound, we have
used the assumption that there is at least a gravitationally suppressed
non-nonrenormalizable operator of dimension 5 or 6 that can contribute
to reheating. This assumption sets a bound on the maximum separation
between the reheating temperature and the expansion rate at the end
of inflation in certain cases.

One has to keep in mind that the maximum derived in this paper has
some obvious caveats. First, since we have only considered linear
spectator scenarios, we have not examined what the maximum blue spectral index would be
if we allowed $\delta\chi$ to be of order $\chi_{0}$ . Since we
have imposed $\chi_{0}>H/(2\pi)$ and $\delta\chi$ is at most of
order $H/(2\pi)$, one might think that the current estimate will
stand even after including the $\delta\chi\gtrsim\chi_{0}$ scenarios.
On the other hand, the $n_{I}-1$ of quadratic isocurvature scenarios
(i.e.~scenarios in which the isocurvature perturbations are proportional
to $\Delta_{s}^{2}\propto\langle\delta\chi^{2}\delta\chi^{2}\rangle$)
is twice that of the linear spectator scenario \cite{Chung:2011xd,Chung:2004nh,Liddle:1999pr}.
However, a preliminary investigation shows that this factor of 2 in
the power only gives an enhancement of the form
\begin{equation}
\max[n_{I}-1]\propto\frac{1-\frac{1}{2N_{e}}\ln[k_{\rm max}/k_{{\rm min}}]}{1-\frac{1}{N_{e}}\ln[k_{\rm max}/k_{{\rm min}}]}
\end{equation}
multiplying a difficult to compute suppression (originating from the quantum
nature of the particle production in contrast with the classical VEV
displacements of the linear spectator scenario), resulting in a similar
maximum spectral bound at best. However, given that the dependence
of the relic density and the spectral amplitude with $n_{I}-1$ is
somewhat complicated due to their dependence on the long time mode
evolution \cite{Chung:2011xd}, it would be worthwhile confirming
the quadratic isocurvature estimate more carefully.

Another caveat is that we have assumed a ``standard'' slow-roll,
effectively single-field inflationary scenario with only one reheating
period. Most non-minimal extensions will dilute the VEV energy density
leading to a smaller upper bound. In that sense, most of the
non-minimal extensions are not likely to change this general
picture. Even in the situation in which $\chi_{0}$ makes a phase
transition after inflation (e.g. $\chi_{0}$ goes from $v_{1}$ to
$v_{2}$) such that $\omega_{\chi}$ (now proportional to $v_{2}^{2}$)
is generated after inflation (thereby evading the inflationary
dilution), since it is really $\delta\chi$ that is diluting during
inflation (even though we have been rewriting it as
$\omega_{\chi}\delta\chi/|\chi_{0}-v_{1}|$ being constant during
inflation), this does not help us to evade the bound.

Finally, we have assumed a sampling of inflationary space characterized
by $\{\epsilon_{i},H_{i},t_{e}\}$, while there are infinitely more
ways to tune the inflationary models. On the other hand, even the
addition of $\epsilon_{i}$ (versus a non-evolving scenario of $H(t)$
during inflation) produced only about a 10\% change in $n_{I}-1$.
Hence, we believe this limitation of sampling is not severely restrictive.

It is indeed intriguing that future cosmological inhomogeneity measurements
of $n_{I}\gtrsim2.4$ may uncover the following new features of dark matter: (i)
dark matter had to have a time dependence in its mass in its evolution
history in the context of an inflationary universe, and (ii) dark matter
mass was of order of the expansion rate during inflation. From our
current model building tool-kit, arguably the most appealing
picture that would emerge is that the dark matter is a field possessing
a fundamental shift symmetry just like the axion.
\begin{acknowledgments}
We thank Lisa Everett for comments on this work.  This work was
supported in part by the DOE through grant DE-FG02-95ER40896.  This
work was also supported in part by the Kavli Institute for
Cosmological Physics at the University of Chicago through grant NSF
PHY-1125897 and an endowment from the Kavli Foundation and its founder
Fred Kavli.
\end{acknowledgments}
\appendix

\section{\label{sec:Background-Sol}Background Solution}

For this section, we set the time at which the observable longest
wavelength mode leaves the horizon to be time $t_{k_{i}}=0$. We can
model a very large class of slow-roll inflationary models with the
Hubble expansion rate function parameterized (with three constants
$\{\epsilon_{k_i}, H_{i}, t_{e}\}$, where $t_{e}$ approximately
replaces $\eta_{V}$ in the usual slow-roll parameterization scheme)
as 
\begin{equation}
H\approx\left\{ \begin{array}{ccc}
H_{i}(1-\epsilon_{k_i}H_{i}t) &  & 0<t<t_{e}\\
\frac{H_{i}(1-\epsilon_{k_i}H_{i}t_{e})}{1+\frac{3}{2}(t-t_{e})H_{i}(1-H_{i}\epsilon_{k_i}t_{e})} &  & t_{e}<t<t_{RH}
\end{array}\right \}, \label{eq:generalfittingfunction-1}
\end{equation}
in which $\Delta t\equiv t-t_{i}$ and $t_{RH}$ is the time of reheating.
This ansatz accurately (at the order of 10\% level) both quadratic
inflation and hybrid inflation. Note also that as long as the number
of efolds is fewer than
\begin{equation}
N_{max}\equiv\frac{1}{2\epsilon_{k_{i}}}\approx\frac{4\pi^{2}M_{p}^{2}\Delta_{\zeta}^{2}(k_{i})}{H_{i}^{2}},
\end{equation}
the quantity $H$ will never go negative.

After the end of inflation, the solution of the field evolution equation 
\begin{equation}
\ddot{\chi}_{0}(t)+3H\dot{\chi}_{0}(t)+m_{\chi}^{2}\chi_{0}(t)=0\label{eq:EOM}
\end{equation}
takes the simple form
\begin{equation}
\chi_{0}(t)=\frac{e_{1}\cos(m_{\chi}\Delta t)+e_{2}\sin(m\Delta t)}{1-\frac{3}{2}H_{i}\Delta t(\epsilon_{k_i}H_{i}t_{e}-1)}
\end{equation}
where $\Delta t=t-t_{e}$. 

We could in principle solve the equation of motion (Eq.~(\ref{eq:EOM}))
exactly in this class of models in terms of hypergeometric functions
and Hermite polynomials
\begin{equation}
\chi_{0}=C_{1}\, H_{\frac{m^{2}}{3\epsilon_{k_i}H_{i}^{2}}}\left(-\sqrt{\frac{3}{2\epsilon_{k_i}}}+\sqrt{\frac{3}{2}\epsilon_{k_i}}H_{i}t\right)+C_{2}\,_{1}F_{1}\left(\frac{-m^{2}}{6\epsilon_{k_i}H_{i}^{2}};\frac{1}{2};\left(\sqrt{\frac{3}{2\epsilon_{k_i}}}-\sqrt{\frac{3}{2}\epsilon_{k_i}}H_{i}t\right)^{2}\right).
\end{equation}
However, because $\epsilon_{k_{i}}$ is small, these special functions
must be evaluated in exponentially large and small numerical regions
and added together. Such a route seems numerically unstable, in addition
to being opaque. In practice, it is easier to handle numerically the solution to
the equation of motion subject to the boundary condition
\begin{equation}
\dot{\chi}_{0}(0)=-\left(\frac{3}{2}-\nu_{i}\right)H_{i}\chi_{0}(0),
\end{equation}
which embodies the assumptions that the spectral index is of order
unity and the field is rolling in a slow-roll fashion, initially. 

We can match the solution before and after the end of inflation to
write the solution after the end of inflation as
\begin{equation}
\chi_{0}(t)=K_{1}\left[\frac{H(t)}{H(t_{e})}\right]\cos(m\Delta t+K_{2})
\end{equation}
\begin{equation}
K_{1}\equiv\sqrt{\mathcal{A}}\chi_{0}(0)\exp\left[-\frac{1}{2}(3-2\nu_{k_{i}})H_{i}t_{e}\right],
\end{equation}
in which $K_{2}$ is a phase. The amplitude is given by
\begin{equation}
\sqrt{\mathcal{A}}=\frac{\chi_{0}(t_{e})e^{\frac{1}{2}(3-2\nu_{i})H_{i}t_{e}}}{\chi_{0}(0)}\sqrt{1+\frac{\left[1-\epsilon_{k_i}H_{i}t_{e}+\frac{2}{3}\dot{\chi}_{0}(t_{e})/\left(H_{i}\chi_{0}(t_{e})\right)\right]^{2}}{1-\frac{4}{9}\nu_{i}^{2}}},
\end{equation}
in which we note that $\chi_{0}(t_{e})\exp\left[\frac{1}{2}(3-2\nu_{i})H_{i}t_{e}\right]$
is the initial value $\chi_{0}(0)$ for $\epsilon_{k_i}=0$. Hence, it is more convenient numerically to solve for $\chi_{0}(t)\exp\left[\frac{1}{2}(3-2\nu_{i})H_{i}t\right]$
than $\chi_{0}(t)$. The exponential suppression of $\chi_{0}(t_{e})\exp\left[\frac{1}{2}(3-2\nu_{i})H_{i}t_{e}\right]/\chi_{0}(0)$
still occurs when $9-4m^{2}/H_{i}^{2}/(1-\epsilon_{k_i}H_{i}t_{e})^{2}<0$.
In this notation, the dark matter fraction $\omega_{\chi}$ is
\begin{equation}
\omega_{\chi}=\frac{K_{1}^{2}}{M_{p}^{2}}\frac{m^{2}}{H^{2}(t_{e})}\frac{T_{RH}}{T_{eq}}R.
\end{equation}
where $R$ is defined in Eq.~\ref{eq:Rdef}.

\bibliographystyle{JHEP}
\bibliography{ref,misc,ConsistencyRelation,Curvaton,Non-Gaussianity,Inflation_general,wimpzilla,deltaN,gravitino_isocurvature,blue_isocurvature,string-reheating}

\end{document}